\documentclass[aps,prl,showpacs,twocolumn,nofootinbib]{revtex4}

\usepackage{amsmath,graphicx}

\begin{document}

\title{A Possible Solution to the $B\to \pi\pi$ Puzzle \\ Using the Principle of Maximum Conformality}

\author{Cong-Feng Qiao}
\email{qiaocf@ucas.ac.cn}
\address{School of Physics, University of Chinese Academy of
Sciences, Beijing 100049, P.R. China}
\address{CAS Center for Excellence in Particle Physics,
Institute of High Energy Physics, Chinese Academy of Sciences, Beijing 100049, P.R. China}
\author{Rui-Lin Zhu}
\email{zhuruilin09@mails.ucas.ac.cn}
\address{School of Physics, University of Chinese Academy of
Sciences, Beijing 100049, P.R. China}
\address{CAS Center for Excellence in Particle Physics,
Institute of High Energy Physics, Chinese Academy of Sciences, Beijing 100049, P.R. China}

\author{Xing-Gang Wu}
\email{wuxg@cqu.edu.cn}
\address{Department of Physics, Chongqing University, Chongqing 401331, P.R. China}

\author{Stanley J. Brodsky}
\email{sjbth@slac.stanford.edu}
\affiliation{SLAC National Accelerator Laboratory, Stanford University, Stanford, CA 94309, USA}

\date{\today}

\begin{abstract}

The measured $B_d \to \pi^0\pi^0$ branching fraction deviates significantly from conventional QCD predictions, a puzzle which has persisted for more than 10 years. This may be a hint of new physics beyond the Standard Model; however, as we shall show in this paper, the pQCD prediction is highly sensitive to the choice of the renormalization scales which enter the decay amplitude. In the present paper, we show that the renormalization scale uncertainties for $B\to \pi\pi$ can be greatly reduced by applying the Principle of Maximum Conformality (PMC), and more precise predictions for CP-averaged branching ratios ${\cal B}(B\to\pi\pi)$ can be achieved. Combining the errors in quadrature, we obtain ${\cal B}(B_{d}\to \pi^0\pi^0)|_{\rm PMC} = \left(0.98^{+0.44}_{-0.31}\right) \times10^{-6}$ by using the light-front holographic low-energy model for the running coupling.  All of the CP-averaged $B\to\pi\pi$ branching fractions predicted by the PMC are consistent with the Particle Data Group average values and the recent Belle data. Thus, the PMC provides a possible solution for the $B_d \to \pi^0\pi^0$ puzzle.



\end{abstract}

\maketitle

$B$-meson hadronic two-body decays contain a wealth of information on the physics underlying charge-parity (CP) violation. Measurements of the $B$-meson two-body branching ratios and their CP asymmetries provide key information on the Cabibbo-Kobayashi-Maskawa (CKM) matrix elements. One challenge which has puzzled the theoretical physics community for more than 10 years is that the measured branching ratio~\cite{Beringer:1900zz, Amhis:2012bh, Aubert:2004aq} for the decay of the B meson to neutral pion pairs $B_d \to \pi^0\pi^0$ is significantly larger than the theoretical predictions based on the QCD factorization approach~\cite{Beneke:2009ek, Burrell:2005hx}, a perturbative QCD approach~\cite{Zhang:2014bsa}, and an Isospin analysis~\cite{Gronau:1990ka}.

Beneke {\it et al.} (BBNS)~\cite{Beneke:1999br} have developed a systematic QCD analysis of $B\to \pi\pi$ based on the factorization of long-distance and short distance dynamics. The BBNS predictions for the branching ratios of $B_d \to \pi^+\pi^-$ and $B^\pm \to \pi^\pm\pi^0$ are consistent with CLEO, BaBar, and Belle data. However, the BBNS prediction for the $B_d \to \pi^0\pi^0$ branching ratio deviates significantly from measurements~\cite{Aubert:2004aq}. There have been suggestions on how to resolve this puzzle and to obtain a consistent explanation of all $B\to\pi\pi$ channels within the same framework. In particular, Beneke {\it et al.}~\cite{Beneke:2005vv} have noted that the one-loop QCD corrections to the color-suppressed hard-spectator scattering amplitude $\alpha_2(\pi\pi)$ could be important, as seen from their calculation of the vertex corrections up to two-loop QCD corrections~\cite{Beneke:2009ek}. However, even after including those higher-order QCD corrections, the discrepancy has remained. The large $K$ factor, $K={\cal B}^{\rm NLO}/{\cal B}^{\rm LO}$, with ${\cal B}^{\rm LO/NLO}$ corresponding to the LO/NLO-terms in the branching ratio ${\cal B}$, implied by the higher-order corrections to the branching ratio of $B_d \to \pi^0\pi^0$, as well as the large renormalization scale uncertainties, have called into question the reliability of pQCD calculations.

In the conventional treatment, the renormalization scale is usually fixed to be the typical momentum flow of the process, or one that eliminates large logarithms in order to make the prediction stable under scale changes. This is simply a ``guess", and the scale ambiguities and scheme-dependence persist at any fixed order. Thus, if one uses conventional scale setting for an $\alpha_s^{n}$-order pQCD prediction, the scale ambiguity is not a $\alpha_s^{n+1}$-order effect, it exists for any known perturbative terms~\cite{PMC3}.

According to {\it renormalization group invariance}, a valid prediction for a physical observable should be independent of theoretical conventions, such as the choices of the renormalization scheme and the renormalization scale. This important principle is satisfied by the Principle of Maximum Conformality (PMC)~\cite{PMC1, PMC2, pmccolloquium}. The running behavior of the coupling constant is controlled via the renormalization group equation. Conversely, the knowledge of the $\{\beta_i\}$-terms in the perturbative series can be used to determine the optimal scale of a particular process; this is the main goal of the PMC. The PMC is a generalization of the well-known Brodsky-Lepage-Mackenzie (BLM) procedure to all orders~\cite{blm}~\footnote{The BLM approach of using the $n_f$-terms as a guide to resum the series through the renormalization group equation of $\alpha_s$ cannot be unambiguously extended to high orders.}. If one fixes the renormalization scale of the pQCD series using the PMC, all non-conformal $\{\beta_i\}$-terms in the perturbative expansion series are resummed into the running coupling, and one obtains a unique, scale-fixed, scheme-independent prediction at any finite order.

In the following, we will apply the PMC procedure to the BBNS analysis with the goal of eliminating the renormalization scale ambiguity and achieving an accurate pQCD prediction which is independent of  theoretical conventions. In fact, as we shall show, the PMC can
provide a solution to the $B\to\pi\pi$ puzzle.

\begin{figure}[tb]
\includegraphics[width=0.45\textwidth]{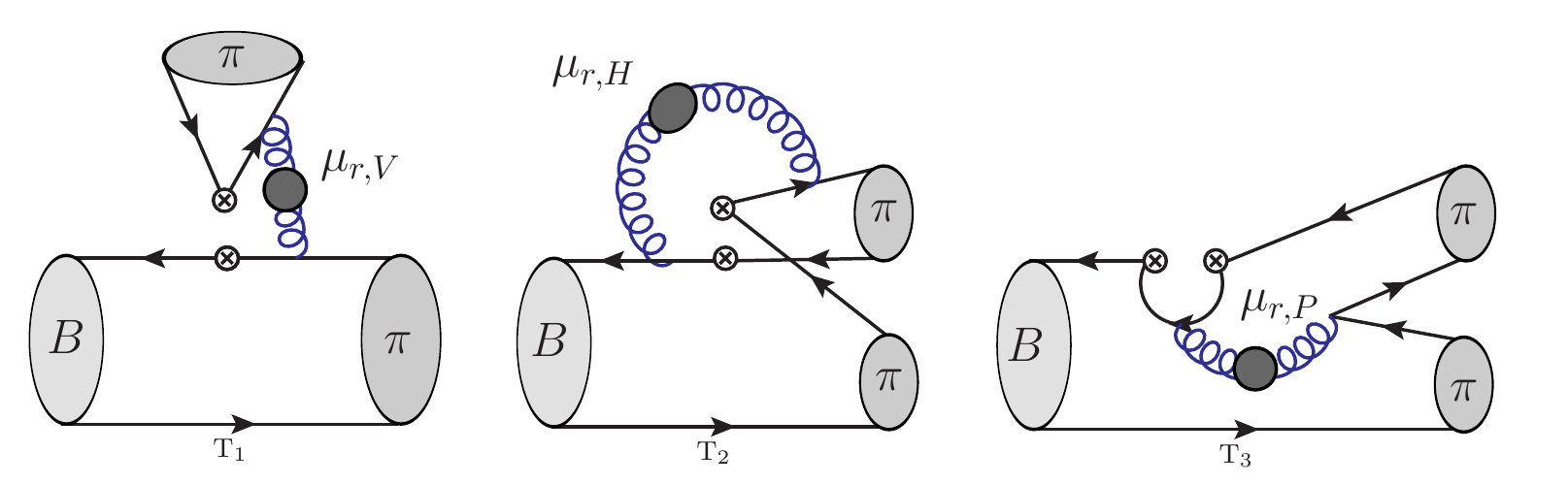}
\caption{Typical Feynman diagrams for the $B\to\pi\pi$ decays, which are sizable and correspond to $\alpha_1$, $\alpha_2$, $\alpha_4$ (or $\alpha_6$), respectively. $\mu_{r,V}$, $\mu_{r,H}$ and $\mu_{r,P}$ are renormalization scales for these diagrams; they are different in general. Other Feynman diagrams can be obtained by shifting one of the gluon endpoints to different quark lines. The vertex ``$\otimes\otimes$'' denotes the insertion of a 4-fermion operator $Q_i$. And the big dot stands for the renormalized gluon propagator whose light-quark loop determines the $\beta_0$-terms and hence the optimal scale for the running behavior of the QCD coupling constant.} \label{feyn}
\end{figure}

The effective weak Hamiltonian~\cite{Buchalla:1996}
\begin{equation}
{\cal H}_{\rm eff} = \frac{G_F}{\sqrt{2}} \sum_{p=u,c} \lambda_p
\left[C_1 Q_1^p+C_2Q_2^p+\sum_{i=3\ldots6} C_i Q_i \right] \ ,
\label{eq:HeffKpp}
\end{equation}
where $\lambda_p=V^*_{pd}V_{pb}$, $Q_i(\mu_f,\mu_r)$ are local four-fermion interaction operators and the $C_i(\mu_f,\mu_r)$ are the corresponding short-distance Wilson coefficients at the renormalization scale $\mu_r$ and the factorization scale $\mu_f \sim m_b$. Applying the QCD factorization, the amplitude for $B\to \pi\pi$ decay, assuming the dominance of valence Fock states for both the $B$ meson and the final-state pions, can be expressed as
\begin{equation}
\langle \pi\pi|{\cal H}_{\rm eff}|\bar{B}\rangle =\frac{G_F}{\sqrt{2}} \sum_{p=u,c}
\lambda_p \langle \pi\pi|{\cal T}_p|\bar{B}\rangle \ ,
\end{equation}
where the right-hand operator that creates the weak transition in the Standard Model is
\begin{widetext}
\begin{eqnarray}
{\cal T}_p&=&\alpha_1^p(\pi\pi)(\bar{u}b)_{V-A}\otimes (\bar{d}u)_{V-A}+
\alpha_2^p(\pi\pi)(\bar{d}b)_{V-A}\otimes (\bar{u}u)_{V-A}
+\alpha_3(\pi\pi)(\bar{d}b)_{V-A}\otimes (\bar{q}q)_{V-A}+\nonumber\\
&&\alpha_4^p(\pi\pi)(\bar{q}b)_{V-A}\otimes (\bar{d}q)_{V-A}+
\alpha_5(\pi\pi)(\bar{d}b)_{V-A}\otimes (\bar{q}q)_{V+A}+
\alpha_6^p(\pi\pi)(-2)(\bar{q}b)_{S-P}\otimes (\bar{d}q)_{S+P}\,.
\end{eqnarray}
\end{widetext}
A summation over $q=u,d$ is implied in this equation, and the required currents are $(\bar{q} q^\prime)_{V\pm A}=\bar{q} \gamma^\mu (1\pm \gamma_5)q^\prime$ and $(\bar{q} q^\prime)_{S\pm P} =\bar{q} (1\pm \gamma_5)q^\prime$. The relations among the Wilson coefficients $C_i$ and $\alpha^{(p)}_j$ can be found in Ref.~\cite{Beneke:2001}. The branching ratio for $B\to \pi\pi$ is given by $ {\cal B}(\bar{B}\to \pi\pi)=\tau_B|{\cal A}(\bar{B}\to \pi\pi)|^2 S/(16\pi m_B)$,
where the symmetry parameter $S=1/2!$ for $\pi^0\pi^0$, and $S=1$ for $\pi^+\pi^-$ or $\pi^\pm\pi^0$, respectively.

Typical Feynman diagrams which provide non-zero contributions to the $B\to\pi\pi$ decays and correspond to $\alpha_1$, $\alpha_2$, $\alpha_4$ and $\alpha_6$, respectively, are illustrated in Fig.(\ref{feyn}). The resulting amplitudes under the $\overline{\rm MS}$-scheme for $B\to \pi\pi$ can be written as~\cite{Beneke:1999br}
\begin{widetext}
\begin{eqnarray}\label{abpm}
&&{\cal A}(\bar B_0\to\pi^+\pi^-) = i\frac{G_F}{\sqrt{2}}m^2_B
f_+^{B \to \pi}(0)f_\pi |\lambda_c|\{R_b e^{-i\gamma} [\alpha^u_1
+\alpha^u_4+\alpha^u_6r_\chi] -[\alpha^c_4+\alpha^c_6r_\chi]\}\,,
\nonumber\\
&&{\cal A}(\bar B_0\to\pi^0\pi^0) = i\frac{G_F}{\sqrt{2}}m^2_B
f_+^{B \to \pi}(0)f_\pi |\lambda_c|\{R_b e^{-i\gamma} [-\alpha^u_2+\alpha^u_4+\alpha^u_6r_\chi]
-[\alpha^c_4+\alpha^c_6r_\chi]\}\,,\nonumber\\
&&{\cal A}(B^-\to\pi^-\pi^0) = i\frac{G_F}{\sqrt{2}}m^2_B
f_+^{B \to \pi}(0)f_\pi |\lambda_c|(R_b/\sqrt{2}) e^{-i\gamma}
[\alpha^u_1+\alpha^u_2]\,,
\end{eqnarray}
\end{widetext}
where $R_b=|V_{ub} V_{ud}^*|/|V_{cb}V_{cd}^*|$, and $\gamma$ is
the $V_{ub}^*$ phase. The coefficient $r_\chi(\mu_r)= 2m_\pi^2/[\bar{m}_b(\mu_r) (\bar{m}_u(\mu_r) +\bar{m}_d(\mu_r))]$, which equals to $1.18$ when setting the scale $\mu_r=m_b$~\cite{Beneke:1999br}. Here $f_\pi (f_B)$ is the pion ($B$-meson) decay constant, and $f_+^{B \to \pi}(0)$ is the $B\to \pi$ transition form factor at the zero momentum transfer. The CP conjugate amplitudes are obtained from the above by replacing $e^{-i\gamma}$ to $e^{+i\gamma}$. The topological tree amplitude $\alpha_1$ expresses the contribution when the final $(\bar{u}d)$-pair (produced from the virtual $W^-$) forms the pion directly. The tree amplitude $\alpha_2$ expresses the contribution obtained when the final $(\bar{u}d)$-pair from $W^-$ separates and one of them forms a pion by coalescing with the spectator quark. The amplitudes $\alpha_i$ (i=3 to 6) are topological penguin amplitudes. Note that when the spectator quark combines with one of the quarks from $W^-$ to form a pion, a color-suppressed factor $\sim 1/N_c$ emerges. Thus, the amplitude $\alpha_1$ provides the dominant contributions relative to the color-suppressed $\alpha_{2,4,6}$. However this color suppression can effectively disappear when one includes higher-order gluonic interactions to $\alpha_{2,4,6}$; their contributions thus can be sizable. At present, consistent pQCD calculations of the tree amplitudes $\alpha_{1,2}$ and their vertex corrections have been evaluated with two-loop QCD corrections. The one-loop QCD correction to the hard spectator scattering interaction has been done by Ref.\cite{Beneke:2009ek}. All of them are up to ${\cal O}(\alpha_s^2)$ level.

We rewrite the contributions in the following convenient form:
\begin{widetext}
\begin{eqnarray}
\alpha^p_{1} &=& C_1(\mu_f,\mu^{\rm init}_{r,V})+ \frac{1}{N_c} \Bigg[ C_F C_2(\mu_f,\mu^{\rm init}_{r,V}) \left\{ \frac{1}{C_F}+\frac{\alpha_s(\mu^{\rm init}_{r,V})} {4\pi} V_{1}(\mu_f, \mu^{\rm init}_{r,V}) + \left( \frac{\alpha_s(\mu^{\rm init}_{r,V})}{4\pi} \right)^2 \beta_0 \tilde{V}_{1}(\mu_f,\mu^{\rm init}_{r,V}) \right\} \nonumber \\
&& +\left( \frac{\alpha_s(\mu^{\rm init}_{r,V})}{4\pi} \right)^2 V_2(\mu_f,\mu^{\rm init}_{r,V})  + \frac{4 C_F C_2(\mu_f,\mu^{\rm init}_{r,H})\pi^2}{N_c} \left\{ \frac{\alpha_s(\mu^{\rm init}_{r,H})}{4\pi}\, H_{1}(\mu_f,\mu^{\rm init}_{r,H}) +  \left( \frac{\alpha_s(\mu^{\rm init}_{r,H})}{4\pi} \right)^2 \beta_0 \tilde{H}_{1}(\mu_f,\mu^{\rm init}_{r,H}) \right\} \nonumber\\
&& +\left( \frac{\alpha_s(\mu^{\rm init}_{r,H})}{4\pi} \right)^2 H_2(\mu_f, \mu^{\rm init}_{r,H})\Bigg], \\
\alpha^p_{2} &=& C_2(\mu_f,\mu^{\rm init}_{r,V})+ \frac{1}{N_c} \Bigg[  C_F C_1(\mu_f,\mu^{\rm init}_{r,V}) \left\{ \frac{1}{C_F}+\frac{\alpha_s(\mu^{\rm init}_{r,V})}{4\pi} V_{1}(\mu_f, \mu^{\rm init}_{r,V}) + \left( \frac{\alpha_s(\mu^{\rm init}_{r,V})}{4\pi} \right)^2 \beta_0 \tilde{V}_{1} (\mu_f,\mu^{\rm init}_{r,V})\right\} \nonumber \\
&& +\left( \frac{\alpha_s(\mu^{\rm init}_{r,V})}{4\pi} \right)^2 V_3(\mu_f,\mu^{\rm init}_{r,V})  + \frac{4 C_F C_1(\mu_f,\mu^{\rm init}_{r,H})\pi^2}{N_c} \left\{ \frac{\alpha_s(\mu^{\rm init}_{r,H})} {4\pi} H_{1}(\mu_f,\mu^{\rm init}_{r,H}) +  \left( \frac{\alpha_s(\mu^{\rm init}_{r,H})}{4\pi} \right)^2 \beta_0 \tilde{H}_{1}(\mu_f,\mu^{\rm init}_{r,H}) \right\} \nonumber\\
&& +\left( \frac{\alpha_s(\mu^{\rm init}_{r,V})}{4\pi} \right)^2 H_3(\mu_f,\mu^{\rm init}_{r,H})\Bigg].
\end{eqnarray}
The penguin diagrams provide small contributions to the amplitudes, which are
\begin{eqnarray}
\alpha^p_{4} &=& C_4(\mu_f,\mu^{\rm init}_{r,V})+ \frac{C_3(\mu_f,\mu^{\rm init}_{r,V})}{N_c}
\Bigg[1+ \frac{\alpha_s(\mu^{\rm init}_{r,V})}{4\pi} C_F  V_{1}(\mu_f,\mu^{\rm init}_{r,V})\,+\frac{\alpha_s(\mu^{\rm init}_{r,V})}{4\pi} \frac{C_F}{N_c} P^p_{\pi,2}(\mu_f,\mu^{\rm init}_{r,V})\Bigg] \nonumber\\
&& + \frac{4 C_3(\mu_f,\mu^{\rm init}_{r,H}) C_F \pi^2}{N^2_c} \frac{\alpha_s(\mu^{\rm init}_{r,H})}{4\pi}\, H_{1}(\mu_f,\mu^{\rm init}_{r,H}), \\
\alpha^p_{6} &=& C_6(\mu_f,\mu^{\rm init}_{r,V}) + \frac{C_5(\mu_f,\mu^{\rm init}_{r,V})}{N_c} \Bigg[1+ \frac{\alpha_s(\mu^{\rm init}_{r,V})}{4\pi}
C_F (-6) \, +\frac{\alpha_s(\mu^{\rm init}_{r,V})}{4\pi} \frac{C_F}{N_c}  P^p_{\pi,3}(\mu_f,\mu^{\rm init}_{r,V})\Bigg]. \label{amp2}
\end{eqnarray}
\end{widetext}
In these equations, the factorization scale dependence and the renormalization scale dependence are explicitly written in the Wilson coefficients and the functions $V_1$, $\tilde{V}_1$, $V_2$, $V_3$, $H_1$, $\tilde{H}_1$, $H_2$, $H_3$, $P^p_{\pi,2}$ and $P^p_{\pi,3}$, where $\mu^{\rm init}_{r,H}$ and $\mu^{\rm init}_{r,V}$ stand for the initial choice of renormalization scales. The corresponding expressions for the functions with explicit renormalization and factorization scale dependence can be found in Eqs. (16), (19), (26), and (30) of Ref.~\cite{Burrell:2005hx}. Here $\beta_0=(11N_c-2n_f)/3$, $V_i$ ($\tilde{V}_i$) denotes the vertex corrections, and $H_i$ ($\tilde{H}_i$) denotes the hard spectator scattering contributions. The $\beta_0$-independent term $V_{2(3)}$ and $H_{2(3)}$ can be obtained in Eqs.~(42-47) of Ref.~\cite{Beneke:2009ek} by Beneke, Huber and Li. The Wilson coefficients are contained implicitly in the terms $V_{2(3)}$ and $H_{2(3)}$. The initial scales are set to $\mu^{\rm init}_{r,P}=\mu^{\rm init}_{r,V}$. The quantity $P^p_{\pi,n}$ refers to the contribution from the pion twist-$n$ light-cone distribution amplitude, the expressions of which can be found in Eqs. (49) and (54) of Ref.~\cite{Beneke:2001}. In the calculation both twist-2 and twist-3 terms are taken into consideration. Note that the Wilson coefficients $C_1$ and $C_2$ are different from the definition of Ref.~\cite{Buchalla:1996}, where the labels 1 and 2 are interchanged.

In order to apply the PMC, we have divided the amplitudes into $\beta_0$-dependent nonconformal and $\beta_0$-independent conformal parts, respectively. There are two typical momentum flows for the process; thus, we have assigned two arbitrary initial scales
$\mu^{\rm init}_{r,V}$ and $\mu^{\rm init}_{r,H}$ for the vertex contributions and hard spectator scattering contributions. In the case of conventional scale setting, the scales are fixed to be their typical momentum transfers, i.e. $\mu_{r,V} \equiv \mu^{\rm init}_{r,V}\sim m_b$ and $\mu_{r,H} \equiv\mu^{\rm init}_{r,H}\sim \sqrt{\Lambda_{\rm QCD} m_b} $.

After applying the standard PMC procedures, all non-conformal $\beta_0$-terms are resummed into the strong running coupling, and the amplitudes become
\begin{widetext}
\begin{eqnarray}
\alpha^{p,{\rm PMC}}_{1} &=& C_1(\mu_f,\mu^{\rm init}_{r,V}) + \frac{1}{N_c} \Bigg[ C_2(\mu_f,\mu^{\rm init}_{r,V}) + C_2(\mu_f,\mu^{\rm init}_{r,V}) C_F  \frac{\alpha_s(Q^V_1)}{4\pi}\,V_{1}(\mu_f,\mu^{\rm init}_{r,V}) + \left( \frac{\alpha_s(Q^V_1)}{4\pi} \right)^2 V_{2}^{\prime}(\mu_f,\mu^{\rm init}_{r,V})  \nonumber\\
&& + \frac{4 C_2(\mu_f,\mu^{\rm init}_{r,H}) C_F \pi^2}{N_c}  \frac{\alpha_s(Q^H_1)}{4\pi}\,
H_{1}(\mu_f,\mu^{\rm init}_{r,H}) +  \left( \frac{\alpha_s(Q^H_1)}{4\pi} \right)^2 H_{2}(\mu_f,\mu^{\rm init}_{r,H})  \Bigg],\label{ampPMC1}
\end{eqnarray}
\begin{eqnarray}
\alpha^{p,{\rm PMC}}_{2} &=& C_2(\mu_f,\mu^{\rm init}_{r,V}) + \frac{1}{N_c} \Bigg[ C_1(\mu_f,\mu^{\rm init}_{r,V}) + C_1(\mu_f,\mu^{\rm init}_{r,V}) C_F  \frac{\alpha_s(Q^V_1)}{4\pi} \,V_{1}(\mu_f,\mu^{\rm init}_{r,V}) + \left( \frac{\alpha_s(Q^V_1)}{4\pi} \right)^2 V_{3}^\prime(\mu_f,\mu^{\rm init}_{r,V}) \nonumber\\
&& + \frac{4 C_1(\mu_f,\mu^{\rm init}_{r,H}) C_F \pi^2}{N_c}  \frac{\alpha_s(Q^H_1)}{4\pi}\, H_{1}(\mu_f,\mu^{\rm init}_{r,H}) +  \left(\frac{\alpha_s(Q^H_1)}{4\pi} \right)^2  H_{3}(\mu_f,\mu^{\rm init}_{r,H}) \Bigg], \label{ampPMC2}
\end{eqnarray}
\begin{eqnarray}
\alpha^{p,{\rm PMC}}_{4} &=& C_4(\mu_f,\mu^{\rm init}_{r,V}) + \frac{C_3(\mu_f,\mu^{\rm init}_{r,V})}{N_c} \Bigg[1+ \frac{\alpha_s(Q^V_1)}{4\pi} C_F V_{1}(\mu_f,\mu^{\rm init}_{r,V}) \,+\frac{\alpha_s(Q^V_1)}{4\pi} \frac{C_F}{N_c} P^p_{\pi,2}(\mu_f,\mu^{\rm init}_{r,V})\Bigg]  \nonumber\\
&&  +\frac{4 C_3(\mu_f,\mu^{\rm init}_{r,H}) C_F \pi^2}{N^2_c}  \frac{\alpha_s(Q^H_1)}{4\pi}\,
H_{1}(\mu_f,\mu^{\rm init}_{r,H}), \label{ampPMC3}\\
\alpha^{p,{\rm PMC}}_{6} &=& C_6(\mu_f,\mu^{\rm init}_{r,V}) + \frac{C_5(\mu_f,\mu^{\rm init}_{r,V})}{N_c} \Bigg[1+ \frac{\alpha_s(Q^V_1)}{4\pi} C_F (-6) \,  +\frac{\alpha_s(Q^V_1)}{4\pi} \frac{C_F}{N_c}  P^p_{\pi,3}(\mu_f,\mu^{\rm init}_{r,V})\Bigg] \,,\label{ampPMC4}
\end{eqnarray}
\end{widetext}
where
\begin{eqnarray}
Q^V_1 &=& \mu^{\rm init}_{r,V}\,\mathrm{exp} \left[-\frac{\tilde{V}_1(\mu_f,\mu^{\rm init}_{r,V})}{2V_1(\mu_f,\mu^{\rm init}_{r,V})}\right], \\
Q^H_1 &=& \mu^{\rm init}_{r,H}\,\mathrm{exp} \left[-\frac{\tilde{H}_1(\mu_f,\mu^{\rm init}_{r,H})}{2H_1(\mu_f,\mu^{\rm init}_{r,H})}\right]
\end{eqnarray}
denote the separate PMC scales for the vertex contribution and the hard spectator scattering contribution, respectively. For the penguin amplitude, there is no $\beta$-terms to determine its PMC scale, we take it as $Q^V_1$, the same as the scale of the vertex amplitude, since both types of diagrams have similar space-like momentum transfers. There is a residual scale dependence due to unknown higher-order $\{\beta_i\}$-terms, which however is highly suppressed~\cite{PMC1,PMC2}.  Both $V_1$ and $\tilde{V}_1$ have an imaginary part. We use the real part to set the PMC scale $Q^V_1$. Thus the function $V^\prime_{2(3)}$ has the same expression of $V_{2(3)}$ except for a non-resummed $\beta_0$-related imaginary part, namely $V_3^\prime(\mu_f,\mu^{\rm init}_{r,V})=V_2(\mu_f,\mu^{\rm init}_{r,V}) +C_F C_2(\mu_f,\mu^{\rm init}_{r,V})\beta_0 {\rm Im} \tilde{V}_1(\mu_f,\mu^{\rm init}_{r,V})$ and $V_2^\prime(\mu_f,\mu^{\rm init}_{r,V})=V_3(\mu_f,\mu^{\rm init}_{r,V})+C_F C_1(\mu_f,\mu^{\rm init}_{r,V})\beta_0 {\rm Im} \tilde{V}_1(\mu_f,\mu^{\rm init}_{r,V})$. The values of the resulting PMC scales are $Q^V_{1}\simeq1.59$ {GeV} and $Q^H_{1}\simeq 0.75$ {GeV}; they are nearly independent of the initial scales $\mu^{\rm init}_{r,V}$ and $\mu^{\rm init}_{r,H}$. One should note that the largest uncertainty of $Q^H_{1}$ comes from the chiral enhancement parameter $r_\chi$, which is implicit in $H_1$ and $\tilde{H}_1$. If the value of $r_\chi$ goes up to $1.42$~\cite{Burrell:2005hx}, the PMC scale $Q^H_{1}$ increases to $0.90$ GeV.

A major problem for the present process is that the PMC scale $Q^H_{1}$ is close to $\Lambda_{\rm QCD}$ in the $\overline {\rm MS}$ scheme.  To avoid this low-scale problem, we have utilized commensurate scale relations (CSR)~\cite{csr,csr2} to transform the ${\overline {MS}}$ running coupling to an effective charge defined from a measured physical process. In particular the coupling $\alpha^{\rm g1}_s(Q)$
defined from the Bjorken sum rule is very well measured. To be consistent with the present treatment of $B\to\pi\pi$, we have adopted the leading-order CSR, which gives $\alpha^{\overline{\rm MS}}_s(0.75 {\rm GeV}) = \alpha^{\rm g1}_s(2.04 {\rm GeV})$~\footnote{It is noted that by using the known next-to-leading order CSR, the final branching ratios are altered by less than $\pm5\%$. }. Furthermore, we have adopted the light-front holography model proposed in Ref.~\cite{st} to obtain an estimate of $\alpha^{\rm g1}_s(Q)$. A recent comparison of  the light-front holographic prediction for $\alpha^{\rm g1}_s(Q)$ with  JLAB data can be found in Ref.\cite{g1new}.  This nonperturbative approach is based on the light-front holographic
mapping of classical gravity in anti-de Sitter space, modified by a positive-sign dilaton background. It leads to a reasonable nonperturbative effective coupling. The confinement potential and light-front Schr\"odinger equation derived from this approach also accounts well for the spectroscopy and dynamics of light-quark hadrons.
Other input parameters are chosen as~\cite{Beringer:1900zz}: the $B$-meson lifetime $\tau_{B^+}=1.641{\rm ps}$ and $\tau_{B_d}=1.519{\rm ps}$; $f_B=0.194$ {GeV} and $f_\pi=0.130$ {GeV}; for the CKM parameters, we use $\gamma=68.6^0$, $|V_{cb}|=0.041$, $|V^*_{cd}|=0.230$ and $|V_{ub}|=4.15\times10^{-3}$. The $b$-quark pole mass $m_b=4.8$ {GeV},
and the $c$-quark pole mass $m_c=1.5$ {GeV}. The $n$-th moment of the
$B$ meson's light-front distribution amplitude is adopted
as $\lambda_B=0.20^{+0.04}_{-0.02}$, $\lambda_1=-2.2$ and
$\lambda_2=11$~\cite{Beneke:2009ek}. The second Gegenbauer moment of
the pion leading-twist distribution amplitude is taken as $a_2^\pi=0.2\pm0.1$ and the $B\to \pi$ form factor at zero momentum transfer is taken as $ f_+^{B \to \pi}(0)= 0.25^{+0.03}_{-0.03}$~\cite{Ball:2004ye}, which is estimated by a next-to-leading order light-cone sum rules calculation. By varying $a_2^\pi$, both the form factor $ f_+^{B \to \pi}(0)$ and the branching ratios shall be altered simultaneously, and the form factor $f_+^{B \to \pi}(0)$ dominant the errors to the branching ratios; so, for convenience, we treat the errors caused by $a_2^\pi$ and $f_+^{B \to \pi}(0)$ as a whole and simply call it the $B \to \pi$ form factor error.

As usual, we fix the factorization scale $\mu_f=\mu^{\rm init}_{r,H}$ or $\mu_f=\mu^{\rm init}_{r,V}$, and vary the initial renormalization scale $\mu^{\rm init}_{r,V}\in[1/2 m_b,2m_b]$ and $\mu^{\rm init}_{r,H}\in[1{\rm GeV}, 2{\rm GeV}]$ for analyzing the renormalization scale uncertainty. In general, the factorization and the renormalization scales are different, thus one has to determine the full factorization and renormalization scale dependent expressions for all of the amplitudes; such full-scale dependence can be derived by using Eqs.(\ref{ampPMC1},\ref{ampPMC2},\ref{ampPMC3},\ref{ampPMC4}) via
a general scale translation~\cite{PMC3}.

\begin{table*}
\centering
\caption{Dependence on the renormalization scale of the CP-averaged branching ratio ${\cal B}(B\to \pi\pi)$ (in unit $10^{-6}$) assuming conventional scale setting and PMC scale setting, where three typical (initial) scales are adopted. The first errors are from the $B \to \pi$ form factor and the second errors are from the $B$-meson moment.
\label{result}} \hspace{-2.5mm}
\begin{tabular}{c|c|c|c|c|c|c}
\hline
 & \multicolumn{3}{c|}{Conventional }  &  \multicolumn{3}{c}{PMC }\\
\cline{2-4} \cline{5-7}
$\mu^{\rm init}_{r,V}$ ; $\mu^{\rm init}_{r,H}$ & $ m_b/2$; $1$ GeV &
$m_b$ ; $1.5$ GeV & $2 m_b$; $2$ GeV & $ m_b/2$; $1$ GeV & $ m_b$ ; $1.5$ GeV & $2 m_b$; $2$ GeV \\
\hline $B^-\to \pi^-\pi^0$
 & $5.32^{+1.12+0.21}_{-1.00-0.29}$ & $5.26^{+1.11+0.19}_{-1.00-0.28}$
 & $5.25^{+1.12+0.18}_{-1.01-0.27}$ & $5.89^{+1.84+0.34}_{-1.65-0.50}$
 & $5.89^{+1.84+0.34}_{-1.65-0.50}$  &$5.89^{+1.84+0.34}_{-1.65-0.50}$  \\
$B_d\to \pi^+\pi^-$
 &  $6.10^{+1.72+0.20}_{-1.50-0.13}$ &  $5.93^{+1.65+0.18}_{-1.46-0.13}$
 &  $5.82^{+1.62+0.17}_{-1.41-0.11}$  &  $5.60^{+0.99+0.50}_{-1.57-0.33}$&
 $5.60^{+0.99+0.50}_{-1.57-0.33}$& $5.60^{+0.99+0.50}_{-1.57-0.33}$ \\
$B_d\to \pi^0\pi^0$
 & $0.47^{+0.07+0.07}_{-0.05-0.10}$ & $0.39^{+0.04+0.07}_{-0.03-0.08}$  &
 $0.36^{+0.03+0.06}_{-0.03-0.08}$  & $0.98^{+0.40+0.18}_{-0.03-0.23}$  &
 $0.98^{+0.40+0.18}_{-0.03-0.23}$  & $0.98^{+0.40+0.18}_{-0.03-0.23}$  \\
\hline
\end{tabular}
\end{table*}

\begin{table}[htb]
\centering
\caption{The CP-averaged ${\cal B}(B\to \pi\pi)$ (in units of  $10^{-6}$). The predicted errors are squared averages of those from the $B \to \pi$ form factor, the $B$-meson moment, the chiral enhancement parameter $r_\chi$, and the factorization scale. For the factorization
scale error, we take $\mu_{f,H}=4.8\pm0.8$ {GeV} and  $\mu_{f,V}=1.5\pm0.3$ {GeV}. The PDG and Belle data are presented as a comparison. } \label{result2}
\begin{tabular}{c|c|c|c}
\hline
${\cal B}r$ ($10^{-6}$) & Data & ~~Conv.~~ & ~~PMC~~  \\
\hline
$B^-\to \pi^-\pi^0$  & ~$5.5\pm0.4$~\cite{Beringer:1900zz} ~& ~ $5.26^{+1.13}_{-1.04}$~ & ~ $5.89^{+1.25}_{-1.23}$~     \\
$B_d\to \pi^+\pi^-$  & $5.12\pm0.19$~\cite{Beringer:1900zz} &$5.93^{+1.67}_{-1.47}$  &  $5.60^{+1.19}_{-1.68}$   \\
$B_d\to \pi^0\pi^0$   & $0.90\pm0.12\pm0.10$~\cite{Belle} & $0.39^{+0.09}_{-0.09}$& $0.98^{+0.44}_{-0.31}$  \\
\hline
\end{tabular}
\end{table}

We present our predictions for the CP-averaged $B\to \pi\pi$ in Tables~\ref{result} and \ref{result2}. The CP-conjugate branching ratios are obtained from the CP-conjugate amplitudes following the same procedures. In Table~\ref{result}, we list two main errors from the non-perturbative $B \to \pi$ form factor and the $B$-meson moment;  whereas in Table~\ref{result2}, the errors are the squared averages of those from the $B \to \pi$ form factor, the $B$-meson moment, the chiral enhancement parameter $r_\chi$ and the factorization scale, respectively. An increased branching ratio is observed after PMC scale setting. This indicates that the resummation of the non-conformal series is important. Ref. \cite{Neubert:2002ix} utilizes a similar resummation based on the large $\beta_0$-approximation~\footnote{A detailed comparison of the predictions using the large $\beta_0$-approximation and the PMC can be found in Ref.\cite{Ma:2015dxa}. }; the resulting predictions for the $(B^{\pm}\to \pi^{\pm}\pi^0)$ branching ratio, although not exactly scheme-independent, are found to be numerically consistent with the PMC predictions within errors. If one assumes conventional scale setting, there are large renormalization-scale uncertainties, especially for the color-suppressed topologically-dominated progresses. In contrast, the ambiguity from the choice of the initial renormalization scale is greatly suppressed by using the PMC.

As shown by Table \ref{result}, after applying PMC scale setting, the renormalization scale uncertainty is greatly suppressed as required. The application of the PMC thus removes one of the most important uncertainties in the analysis of $B$ decays, and it provides a sound basis for analyzing higher-twist effects and other possible physics corrections. Table \ref{result2} shows that all the CP-averaged branching ratios of $B\to \pi\pi$ are consistent with the data after PMC scale-setting. By adding the mentioned errors in quadrature, we obtain ${\cal B}(B_{d}\to \pi^0\pi^0)|_{\rm Conv.} = \left(0.39^{+0.09}_{-0.09}\right) \times 10^{-6}$  and ${\cal B}(B_{d}\to \pi^0\pi^0)|_{\rm PMC}= \left(0.98^{+0.44}_{-0.31}\right) \times10^{-6}$, where `Conv.' means calculated using conventional scale setting. After PMC scale-setting, the central value for ${\cal B}(B_{d}\to \pi^0\pi^0)$ is increased by $\sim 100\%$  in comparison with the conventional result $(0.47^{+0.09}_{-0.16} )\times 10^{-6}$. If we had more accurate non-perturbative parameters such as the $B\to\pi$ form factor etc., we could achieve a more precise pQCD prediction. One can define the ratio ${\cal R}_\pi(\pi^-\pi^0)= \Gamma(B^-\to \pi^-\pi^0)/(d\Gamma(B_{d}\to \pi^+ \ell^-\bar{\nu}_\ell)/d q^2|_{q^2=0}$) to cut off the uncertainty from the $B\to\pi$ form factors. In the QCD factorization framework, we have ${\cal R}_\pi(\pi^-\pi^0)= 3\pi^2 f_\pi^2 |V_{ud}|^2|\alpha_1+\alpha_2|^2$, which leads to ${\cal R}_\pi(\pi^-\pi^0)|_{\rm PMC}= 0.87^{+0.08}_{-0.10}$. This is consistent with the heavy flavor averaging group prediction $0.81\pm0.14$~\cite{Amhis:2012bh} within errors.

In summary, we have shown how to use the PMC to eliminate the renormalization scale ambiguity for the QCD running coupling, solving a major problem underlying predictions for $B$-meson decays. The PMC provides a systematic and unambiguous way to set the renormalization scale for QCD processes. The PMC predictions are scheme-independent, as required by renormalization group invariance, and the resulting conformal series avoids the divergent renormalon series. Thus the PMC greatly improves the precision of tests of the Standard Model.

We have applied the PMC with the goal of solving the $B_d \to \pi^0\pi^0$ puzzle. After applying the PMC, the non-conformal $\beta_0$-dependent terms are resummed into the running coupling, and we obtain the optimal scales $Q^V_{1}\simeq1.59$ {GeV} and $Q^H_{1}\simeq 0.75-0.90$ {GeV} for those channels. It is found that the uncertainty of $Q^H_{1}$ come primarily from the chiral enhancement parameter $r_\chi$, which accounts for part of the $\Lambda_{\rm QCD}/m_b$ corrections. The analysis of $\Lambda_{\rm QCD}/m_b$ corrections has been performed in Refs.\cite{masseff1,masseff2}, in which some model-dependent parameters have been introduced with large uncertainties. It has been noted that there are potentially non-perturbative resonance effects that lead to highly suppressed contributions to charm-penguin amplitudes, which however do not invalidate the standard picture of QCD factorization~\cite{masseff3}. As a rough estimate of such uncertainties, we have set $r_\chi=1.42$~\cite{Burrell:2005hx}, which leads to $Q^H_{1}=0.90$ GeV. In comparison with the PMC predictions with $Q^H_{1}= 0.75$ GeV listed in Table~\ref{result2}, such a choice of $r_\chi$ decreases the branching ratio of $B_d\to \pi^0\pi^0$ ($B^-\to \pi^-\pi^0$) by about $10\%$ ($2\%$) and increases the branching ratio of $B_d\to \pi^+\pi^-$ by about $6\%$. This treatment may not exhibit all of the potentially important power-law effects~\footnote{For example, higher Fock states in the B wave function containing charm quark pairs can mediate the decay via a CKM-favored $b\to s c \bar{c}$ tree-level transition. Such intrinsic charm contributions can also be phenomenologically significant~\cite{Brodsky:2001yt}.}, and it is possible that such contributions could yield significant corrections to our present PMC predictions. The uncertainties arising from higher-twist operators is an important theoretical issue which has not been solved.

The PMC results for $B^-\to \pi^-\pi^0$ and $B_d\to \pi^+\pi^-$ are not very different in comparison with traditional predictions, which are already consistent with the data: for $B^-\to \pi^-\pi^0$, the difference is about $10\%$; for $B_d\to \pi^+\pi^-$, the difference is less than $10\%$. However, the situation is quite different for $B_d\to \pi^0\pi^0$,which is dominated by the color-suppressed vertex and power-suppressed penguin diagrams. The difference between the PMC prediction and the traditional prediction is  $\sim100\%$.  However, the PMC prediction agrees with the recent preliminary Belle result ${\cal B}(B_{d}\to \pi^0\pi^0) =(0.90\pm0.12\pm0.10) \times 10^{-6}(6.7\sigma)$~\cite{Belle}. The PMC prediction will become more precise when the nonconformal terms are determined to higher order in the strong coupling $\alpha_s$. Thus, the PMC provides a possible solution for the $B_d \to \pi^0\pi^0$ puzzle.

As a final remark, we have found that the factorization scale uncertainty brings an additional $5\%-10\%$ uncertainty into the pQCD prediction. The factorization scale uncertainty occurs even for a conformal theory; thus the problem of setting the factorization scale reliably at a finite order is unsolved, leading to an additional  systematic uncertainty.  Recently, it has been found that by setting the renormalization scale using the PMC, one substantially suppresses the factorization scale dependence~\cite{facnew}. This again shows the importance of proper renormalization scale-setting. \\

{\bf Acknowledgements:} We thank Guido Bell, Jian-Ming Shen, Alexandre Deur, 
Guy de Teramond and Susan Gardner for helpful discussions. 
This work was supported in part by the Ministry of Science and Technology
of the People's Republic of China under the Grant No.2015CB856703 and 
by the National Natural Science Foundation of 
China under the Grant No.11275280, No.11175249 and No.11375200, the 
Fundamental Research Funds for the Central Universities under the Grant
No.CDJZR305513, and the Department of Energy Contract No.DE-AC02-76SF00515. 
SLAC-PUB-16047.  \\

\end{document}